**Social Media Data Mining With Natural Language Processing on Public Dream Contents**

Howard Hua

Joe Yu

Washington University in St. Louis

University of Michigan

## Abstract

The COVID-19 pandemic has significantly transformed global lifestyles, enforcing physical isolation and accelerating digital adoption for work, education, and social interaction. This study examines the pandemic's impact on mental health by analyzing dream content shared on the Reddit r/Dreams community. With over 374,000 subscribers, this platform offers a rich dataset for exploring subconscious responses to the pandemic. Using statistical methods, we assess shifts in dream positivity, negativity, and neutrality from the pre-pandemic to post-pandemic era. To enhance our analysis, we fine-tuned the LLaMA 3.1-8B model with labeled data, enabling precise sentiment classification of dream content. Our findings aim to uncover patterns in dream content, providing insights into the psychological effects of the pandemic and its influence on subconscious processes. This research highlights the profound changes in mental landscapes and the role of dreams as indicators of public well-being during unprecedented times.

**Social Media Data Mining With Natural Language Processing on Public Dream Contents**

**Introduction**

      The global pandemic brought by COVID-19 has significantly transformed our lives , imposing physical isolation and social activity restrictions (Ingram et al., 2021). These unprecedented changes have led to a substantial change in lifestyle for millions around the world. The pandemic gave popularity to online services like Zoom, Netflix, or Twitter, and hindered the development of many offline activities (Gupta & Singharia, 2021). With governments and global health organizations enforcing social distancing and quarantine policies, individuals had to adapt to a different lifestyle. The closure of offices and schools, along with many public spaces caused a digital transformation during that time period, forcing many workers to work from home and students taking classes remotely (Ingram et al., 2021). This shift, while maintaining productivity and education efficiency, has also introduced adversities like isolation, burnout, or a blurred boundary between work and personal life. The reliance on digital devices during the pandemic matter for work or entertainment also leads to a sedentary lifestyle (Gupta & Singharia, 2021). Moreover, the pandemic has highlighted and exacerbated existing inequalities. Access to technology, safe working conditions, and healthcare have varied significantly, affecting people's ability to cope with changes brought about by the pandemic. Vulnerable populations, mostly the elderly, the newborns, and the ones with pre-existing health conditions, face greater risks of infection (Ahmed, 2020). Overall, the pandemic's physical restrictions and social distancing have profoundly altered lifestyles globally. This shift to a more isolated existence has accelerated digital adoption for work, education, and social interaction, with many other changes in our daily lives, ultimately creating dramatic changes in our lifestyles.

      The COVID-19 pandemic has brought profound changes to both people's physical and mental landscapes back then, and dreams could be an indicator into one's subconscious responses in face of these challenges. While in search of reliable sources of dream contents of the public, the Reddit community came into my attention, which is a popular social media platform nowadays . Specifically, the Reddit r/Dreams server has a total of 374,000 subscribers and is one

of the top 1% populated communities on the Reddit forum . There are also moderators for this server to maintain the quality of the dream posts and delete inappropriate contents (Reddit dreams: Everything about dreams). These qualities make posts in Reddit r/Dreams ideal for data acquisition and further analysis.

In this study, we will delve into the impact of the COVID-19 pandemic on the mental health of the public by analyzing their dreams based on dream posts on the popular online forum—Reddit. Utilizing a statistical approach, our research aims to ascertain whether the emergence of COVID-19 triggered a shift in the contents of dreams. To be specific, we will use statistics to represent the positivity, negativity, and neutrality in each post ranging from the pre-pandemic era to the post-pandemic era. Then we will use mathematical tools to identify patterns and major shifts of positivity, negativity, and neutrality in people's dream posts, which could offer valuable insight into how the global pandemic events affect our psychological well-being and subconscious dreaming.

**Literature Review**

In recent years, the intersection between dreams and mental health had garnered increasing attention, particularly in light of the profound lifestyle alterations brought by the COVID-19 pandemic. As Usher and Durkin (2020) underscore, the global upheaval caused by the pandemic has caused a surge in mental health challenges worldwide. This surge has prompted scholars to delve into the complicated relationship between dream patterns and psychological well-being. As Solomonova and Carr (2019) mentioned in their publication, shifts in dream content may serve as indicators of underlying shifts in mental health states. Building upon this foundation, researchers such as Mellman et al. (2001) and Cvetković et al. (2022) have further explored the nuances of this correlation, shedding light on the potential diagnostic and therapeutic implications of dream analysis in the field of mental health.

Exploring the impact of the COVID-19 pandemic on dreams and mental health had been a focal point of research endeavors. Liu et al. (2022) through their international study on Dream-enactment behaviors during the COVID-19 pandemic substantiated the notion that the pandemic has induced notable changes in dream experiences among individuals globally. Another multi-national research led by Frankl et al. (2021) proved an increase in sleep behavior disorder during the COVID-19 pandemic. These findings revealed significant alterations in dream patterns, highlighting the intricate interplay between the pandemic, dream content, and mental health outcomes. Furthermore, a more detailed field study of dream contents in an Italian town during the COVID-19 lockdown revealed negative emotions like fear, fright, or terror tend to be prevalent when people are experiencing a collective stressful event (Giovanardi, 2021). Their findings collectively underscore the profound impact of the COVID-19 pandemic on the realm of dreams and its implications for mental health.

      In parallel with investigations involving field studies or surveys into dream and mental health during the COVID-19 era, researchers have explored innovative methodologies, including mathematical models, to understand the dynamics of public emotion and its correlation with the pandemic. Li et al. (2020) analyzed public emotion evolution during China's COVID-19 outbreak via social networks. They constructed a dataset from microblogs, tracking emotions and key events from December 2019 to February 2020. Results showed shifting attention from COVID-19 development to other aspects like support and treatment. Notably, public feedback preceded official accounts, and different user groups focused on distinct events during each pandemic phase. Likewise, Nimmi et al. (2021) collected data from social media platforms and used such data to create a deep learning model that analyzes the emotions in the emergency response support system (ERSS) during COVID-19. These studies underscore the potential for broader application of mathematical models and deep learning in large-scale sentiment analysis, offering insights into the complex relationship between COVID-19 and mental health outcomes.

      Previous research applied deep learning or various statistical approaches to COVID-19 sentiment analysis with noticeable results, prompting researchers to implement similar mathematical modeling practices in the analysis of dreams. For instance, Pesonen et al. (2020) explored the impact of the pandemic-induced lockdown on dream content through crowd sourcing, revealing a notable increase in sleep duration alongside heightened frequencies of

awakenings and nightmares with unsupervised computational network and cluster analysis. Additionally, Conte et al. (2022) examined the evolution of dream features across different phases of the pandemic, highlighting significant alterations in dream frequency, length, and affective tone during periods of total and partial lockdown. These studies underscored the complex interplay between dream characteristics and external stressors, a theme echoed in numerous other papers. While there is substantial literature on this topic, few studies have applied deep learning models to analyze COVID-19-era dreams. This gap in the literature serves as the primary motivation for the current research.

## Methodology

### Data Collection and Pre-processing

The dataset input for this study is obtained from the Reddit Application Programming Interface (API) (Reddit dreams: Everything about dreams). After we downloaded the raw posts from Reddit r/Dreams server, we customized the time range from January 1, 2020 to January 1, 2023. Later, with Natural Language Toolkit library (NLTK), python data analysis library (PANDAS) (McKinney, 2010; Bird et al, 2009), we filtered the selected posts using the following techniques:

1. All special characters and newline characters were removed from the dataset.
2. The dataset was changed to all lowercase.
3. Empty entries (deleted posts) were removed from the dataset.
4. Remove all stopwords (i.e. common words such as "a", "an", "the", "and", etc.) and

punctuation in the dataset. Through these pre-processing steps, 98,147 cleaned entries are selected from the 105,000 raw dream posts. None-dream related contents are neglected in this study, as the Reddit r/Dreams server moderators manually supervise the server posts to prevent unrelated scam messages or advertisements from entering the server.

### Pre-trained Llama-3.1 Model

Llama 3.1, an open-source, cutting-edge language model developed by Meta AI, represents a significant leap forward in natural language processing (NLP). Building on the success of its predecessors, Llama 3.1 incorporates advanced training techniques, including increased parameter efficiency and refined data curation strategies (Grattafiori et al., 2024). By leveraging an extensive and diverse dataset, Llama 3.1 excels in a wide array of language understanding tasks. In this study, the Llama 3.1 model is employed and fine-tuned to analyze sentiments in Reddit dream posts.

**Utilizing LoRA in our Training Process**

In our training process, we employ Low-Rank Adaptation (LoRA) to efficiently fine-tune our language model with reduced computational overhead. LoRA allows us to adapt pre-trained models by injecting trainable low-rank matrices into the transformer layers, significantly decreasing the number of parameters needed for fine-tuning (Hu et al., 2021). This approach not only accelerates the training process but also conserves memory, making it feasible to fine-tune large models on limited hardware resources. By leveraging LoRA, we can swiftly adapt the model to specific tasks, such as sentiment analysis in Reddit dream posts, without compromising performance or accuracy.

## Results and Discussion

To effectively leverage the 98,147 data points we collected, we manually cleaned and organized the data to ensure accuracy and reliability. We then introduced the LLaMA-3.1-8B model to fine-tune a sentiment analysis model specifically for COVID-related dream posts. This model was trained to recognize and categorize the emotional content of these posts into six primary sentiment scores: "none-dream related", "joy", "anger", "sadness", "confusion", "neutral", or "fear". As these six sentiment categories are highly representative of social media posts as commonly used among other sentiment analysis tasks (Nimmi et al., 2021). We annotated half of the dataset, which are 49073 entries, our first annotator is a student from Washington University in St. Louis, majoring in computer science and mathematics; our second annotator is a student from University of Michigan, majoring in computer engineering.

To gain insights into the emotional trends over time, we grouped the data points by week. For each week, we calculated the average scores for each sentiment category. The transformer model ensures that these scores always sum up to one for each individual post, allowing for a balanced representation of sentiment. By averaging these scores weekly, we could generate a reflective overview of the emotional landscape during the pandemic.

We then plotted these averaged scores against time to visualize how the emotional content evolved. This approach enabled us to identify patterns and shifts in the collective emotional experience, providing valuable insights into how the pandemic affected people's dreams and emotional states. The visualization helped highlight significant trends and potential correlations with real-world events, offering a deeper understanding of the population's psychological response during this period.

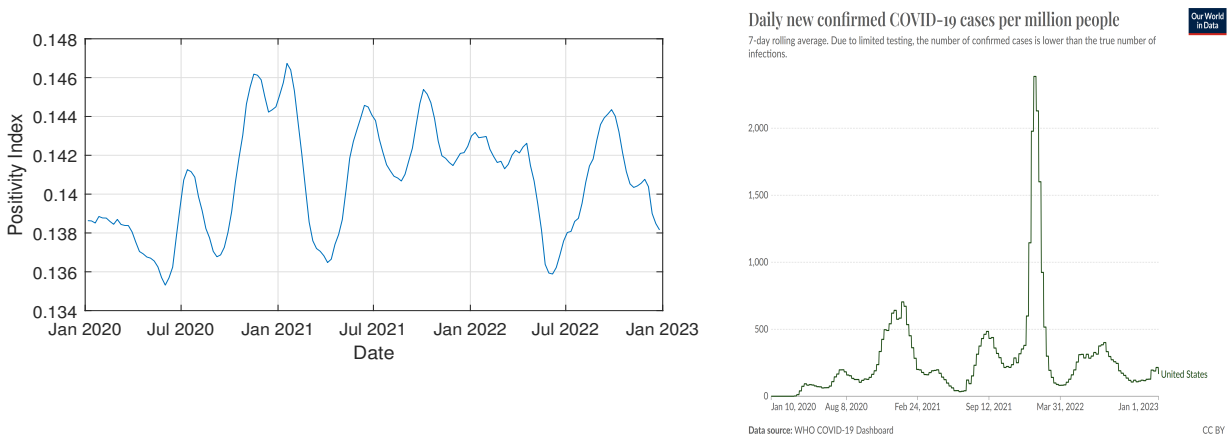

**Figure 1**

*(a) Weekly average positive score in dreams among Reddit dream posts from January 1, 2020 to January 1, 2023 (left); (b) daily new confirmed COVID-19 cases (right).*

In Figure. 1 (a), we observed that the average positive score consistently remains above 0.134 throughout the entire time period, yet never exceeds 0.15. This indicates that positive sentiment accounts for less than 15% of the total emotion, suggesting its rarity compared to the other two sentiments (negative and neutral) in the Reddit dream posts. There are three noticeable drops in the average positive score curve, which occurred around early 2020, late 2021, and late 2022. The first and third drops in the average positive score do not have any close relation with the daily new confirmed COVID-19 cases in the U.S.; however, the second and most significant drop in positive score aligns with the most severe spike of infection cases in the U.S., suggesting that during harsh COVID times, the overall population tends to have less positive dreams when subject to high chances of infection. Supporting this notion, following the late 2021 infections, a second wave of infections came due in late 2022 due to the emergence of the new Omicron variant, leading to an immediate drop in the average positive score curve. But soon after as the infection cases go flat, the positive score soars again.

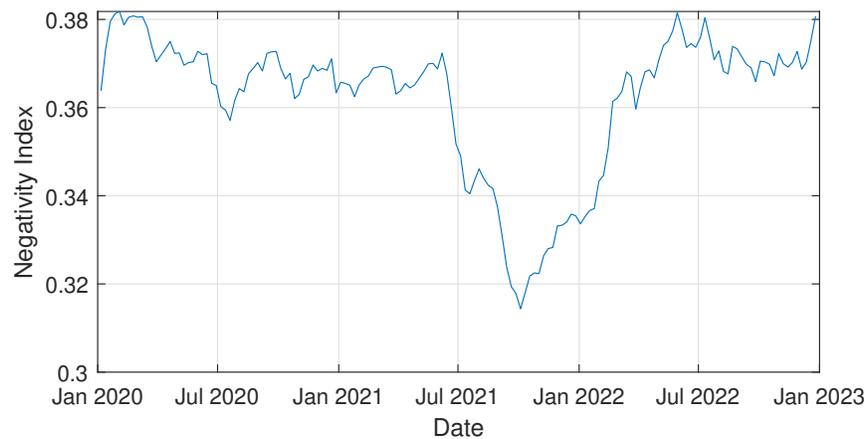

**Figure 2**

*Weekly average negative score in dreams among Reddit dream posts from January 1, 2020 to January 1, 2023.*

Different from the average positive score curve, the average negative score (Figure. 2) maintains a high value ranging from 0.3 to 0.4 between 2020 to 2023. Compared with the 0.15 average positive score, this statistic suggests Reddit users tend to post about their nightmares more often than their sweet dreams on Reddit. The negative score curve maintains a consistent value from 2021 to 2023, except for the late months of 2021.

This connection between dream content and external events is a fascinating area of study and reinforces the concept that dreams can serve as a barometer for societal stressors (Cvetković et al., 2022; Usher & Durkins, 2020; Mellman, 2001). If we set the COVID-19 infection case curve on its side, we could notice the drop in the negative score accord with the drastic drop in infection cases in early 2021. This paralleled movement agrees with Cvetković et al. (2022)'s idea that dreams could be reflective of the external mental state of a population as in this case, the end of a pandemic infection wave brings about a considerable decrease in negative sentiment in the population's dreams, offering a form of indirect proof that the external pressure coming from our daily life diminished temporarily. However, the negative score soon recovers to a normal level over the three months after the major infection wave, indicating the public gradually returning to their normal emotional states.

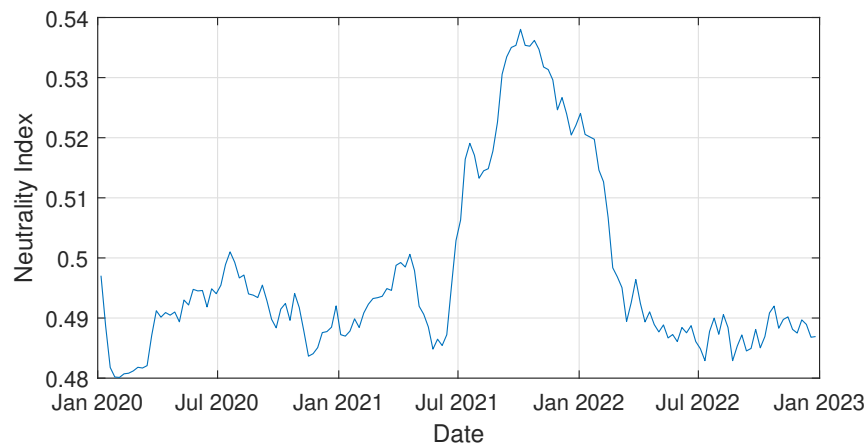

**Figure 3**

*Weekly average neutral score in dreams among Reddit dream posts from January 1, 2020 to January 1, 2023.*

Figure 3 pictures the average neutral score curve between 2020 to 2022. The average neutral score consistently stays above 0.48, with a peak of 0.54, which might be attributed to the nature of the transformer model. If sophisticated dream topics are discussed where no clear signs of positive or negative phrases are mentioned, the model will assign its scores primarily to the "Neutral" class. The neutral score curve shows almost no change between early 2020 to late 2021, except for some minor fluctuations. During the early months of 2022, the average neutral score experienced a surge, while it gently dropped back to the normal standard throughout early to mid-2022. This slow decrease in the neutrality score coincides with the rebound of the negative score and covers the two peaks of the infection case curve between September 2021 to March 2022. These observations suggest that changes in the neutrality score are highly correlated with the alteration of the other sentimental scores, and may reflect broader fluctuations in emotional states among the population, potentially influenced by external factors such as pandemic-related events and public health crises.

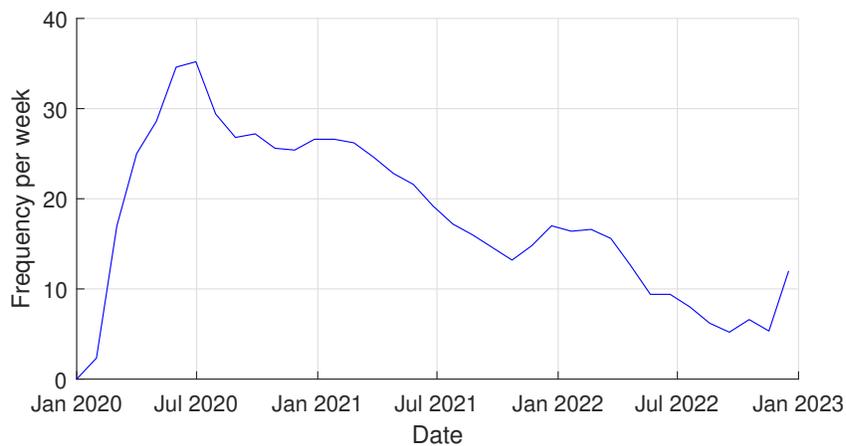

**Figure 4**

*Weekly occurences of COVID-19 related words (i.e. vaccines, COVID, and COVID-19) in Reddit dream posts from January 1, 2020 to January 1, 2023.*

Afterward, we explored the frequency of COVID-19 related phrases such as vaccines, COVID, or COVID-19 (Figure 4). The findings reveal a notable trend: occurrences of COVID-19 related words surged rapidly in the initial months of 2020, yet gradually declined over time. However, notable exceptions include the onset of the first COVID-19 wave around July 2020 and the major Omicron wave in early 2022 (Figure 1. b), during which the frequency of COVID-19 related phrases dramatically increased. The public tends to get used to COVID-19 over time yet new infection waves still have a significant impact on their dream contents.

Together, these scores and common phrases suggest that the COVID-19 pandemic has had a profound effect on the population's mental health, as seen through the lens of dream content on Reddit's r/Dreams server. The correspondences between dream sentiments and real-world events reinforce the notion that our dreams often mirror our waking concerns. As we navigate through and beyond the pandemic, the description of seemingly useless dreams on social platforms continues to serve as a valuable barometer for public mental health.

Finally, we tested the raw performance of the LLaMA 3.1 8B model before proceeding with fine-tuning. The fine-tuning process involved specific configurations to optimize the model for our task. We used the following parameters for the Lora configurations: `lora alpha` was set to 16, `lora dropout` to 0, and `r` to 64. For the training arguments, we configured the number of training epochs of 1.5 and have the per-device train batch size as 3. We used the AdamW optimizer with a learning rate of $2 \times 10^{-4}$ and a weight decay of 0.001.

In figure 5, the curve of the training loss over steps shows an overall downward trend, indicating that the model is learning and improving its predictions. Initially, there's a rapid decrease in loss, suggesting effective early learning. As the training progresses, the curve begins to flatten, indicating the model is approaching convergence. However, there are fluctuations and small spikes throughout the curve, which could be due to noise in the data or the learning rate being slightly high. Towards the end, the loss stabilizes at a lower value, suggesting successful convergence, although some minor oscillations persist.

Figure 6 illustrates the performance of the LLaMA 3.1-8B model in a classification task

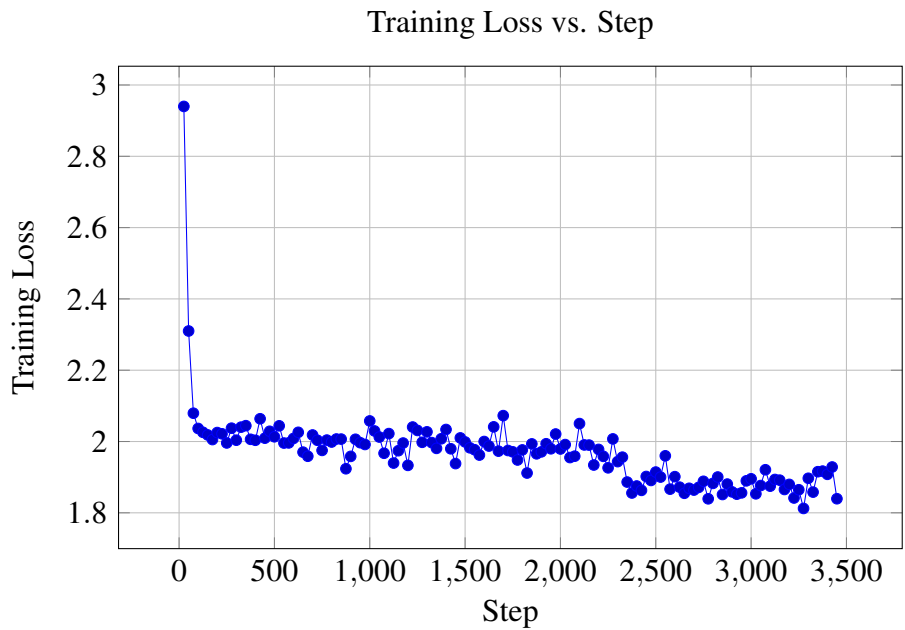

**Figure 5**

*Loss function against steps for Llama-3.1-8B model fine-tuning with SFT trainer.*

when tested against the test dataset, represented by confusion matrices. The left matrix shows the model's performance before fine-tuning, while the right matrix depicts the performance after fine-tuning. The initial model shows a varied distribution of correctly and incorrectly classified instances across different labels. Notably, label 0 has a relatively high number of correct predictions (205), whereas labels 1 "joy" and 2 "anger" show significant mis-classifications. After fine-tuning, there is a noticeable improvement in certain areas, particularly for label 0 "none-dream content", which shows a significant increase in correct predictions (431). However, some labels still exhibit challenges, indicating areas for further optimization. As the overall accuracy increased from 0.395 to 0.419, indicating room for further improvement. However, significant progress was observed in the tags that occur the most frequently in the test data set: "confusion" (label 4) and "fear" (label 6), as these two tags contribute to 67.1% of the test dataset.

　　　　Overall, the fine-tuning process has led to improvements in model accuracy and precision for specific labels, demonstrating the effectiveness of the process in enhancing model performance. However, continued efforts are needed to address remaining classification

| (a) Raw Model | | | (b) Fine-Tuned Model | | |
|---|---|---|---|---|---|
| Label | Accuracy | Confusion Matrix | Label | Accuracy | Confusion Matrix |
| 0 | 0.413 | $\begin{bmatrix} 205 & 60 & 4 \\ 176 & 449 & 7 \\ 50 & 5 & 80 \end{bmatrix}$ | 0 | 0.869 | $\begin{bmatrix} 431 & 3 & 0 \\ 554 & 162 & 1 \\ 105 & 1 & 81 \end{bmatrix}$ |
| 1 | 0.538 | | 1 | 0.194 | |
| 2 | 0.305 | | 2 | 0.309 | |
| 3 | 0.537 | | 3 | 0.571 | |
| 4 | 0.285 | | 4 | 0.482 | |
| 5 | 0.408 | | 5 | 0.000 | |
| 6 | 0.486 | | 6 | 0.232 | |

**Figure 6**

*Comparison of Accuracy and Confusion Matrices before and after fine-tuning the LLaMA 3.1-8B Model: (a) raw model, (b) fine-tuned model. The six labels denote the following emotions: "none-dream content" (0), "joy" (1), "anger" (2), "sadness" (3), "confusion" (4), "neutral" (5), and "fear" (6).*

challenges.

## Conclusion

In conclusion, our study of dream content posted on Reddit from 2020 to 2023 has revealed a distinctive pattern of emotional response to the COVID-19 pandemic. Initially, we focused on three emotions—positive, negative, and neutral—but expanded to train on six emotions, offering deeper insights into the collective mental state during this period of global stress. While this research provides valuable insights into dream content and its reflection of the population's mental status quo, it has several limitations. Notably, while correlations between dream sentiment and pandemic trends are identified, there is no direct evidence that pandemic trends cause specific dream content for individuals. Additionally, the study could be strengthened by using a broader data source from various social platforms worldwide. This research serves as an initial attempt at large-scale dream analysis as a mirror of societal stress, and more advanced training techniques can be applied and will be attempted in future research to further explore the interplay between global events and individual subconscious experiences.

## Acknowledgement

Special thanks to Dr. Gregory for her substantial contributions to drafting this paper, particularly in shaping the literature review and discussion sections.